\documentclass[a4paper,11pt]{article}
\usepackage{pos}

\title{Inflation from Anomalies}

\author*[a,b]{Nick E. Mavromatos}
\author[a]{Panagiotis Dorlis}
\author[a]{Sotirios-Neilos Vlachos}

\affiliation[a]{Physics Division, School of Applied Mathematics and Physical Sciences,
 National Technical University of Athens, Zografou Campus, 9 Iroon Polytechneiou Str., Athens 15780, Greece}

\affiliation[b]{Theoretical Particle Physics and Cosmology Group, Department of Physics, King's College London, Strand, London WC2R 2LS, UK}

\emailAdd{mavroman@mail.ntua.gr}
\emailAdd{psdorlis0@gmail.com}
\emailAdd{sovlacho@gmail.com}

\abstract{We review a string-inspired model of inflation which is a consequence of condensates of chiral gravitational waves (GW) in the primordial Universe, leading in turn to a (approximately) constant condensate of a gravitational anomaly term of Chern-Simons (CS) type, present in the Lagrangian density that describes the dynamics of the very early Universe in the model. We discuss some mechanisms for the production of chiral GW, as well as the role of periodic modulations of the potential of the gravitational axion field, that couples to the CS anomaly term, in ensuring the correct inflationary slow-roll phenomenology of this model.}

\FullConference{Proceedings of the Corfu Summer Institute 2024 "School and Workshops on Elementary Particle Physics and Gravity" (CORFU2024)\
12 - 26 May, and 25 August - 27 September, 2024\\
Corfu, Greece\\}

\begin{document}
\maketitle

\section{Review of the Anomaly-Condensate Inflation}\label{sec{condinfl}}

In \cite{bms3,bms,bms2,ms1,ms2} we have proposed the idea that inflation could arise from condensates of appropriate gravitational Chern-Simons (gCS) anomaly terms due to {\it chiral} (left-right asymmetric) gravitational waves (GW). 
The so-induced inflation is of the Running-Vacuum-Model (RVM) type~\cite{rvm1,rvm2,Lima:2013dmf,SolaPeracaula:2025yco}, in the sense that the condensate is found to be proportional to the quartic power of the Hubble parameter, which dominates the vacuum energy density of the Universe at early eras. In our approach, this non-linear dependence drives the exit from the inflationary phase, and a prolonged reheating period for the Universe, as in the RVM approach~\cite{Lima:2013dmf,Lima:2015mca}, whilst inflation itself is 
driven by the combined role of the gCS condensate and 
the axion field, which couples to the gCS anomaly terms in the Lagrangian.

Detailed computations of the induced condensate, within the weak quantum gravity approach, have been performed in \cite{Dorlis:2024yqw}, going beyond the approximations of \cite{Alexander:2004us,Lyth:2005jf}, which led to
the value of the gCS condensate reduced by a half, as compared to those works. Moreover, the analysis in \cite{Dorlis:2024uei} has demonstrated that, with the exception of the Weyl (symmetric) ordering of quantum-graviton operators, in all other ordering schemes, there are finite (in the presence of the UV cutoff) imaginary parts in the gCS condensate, which amount to a finite lifetime of the inflationary vacuum.

In this section, we shall review the basic features of this scenario, giving emphasis on comparing the field theoretic approach to inflation, via the calculation of the condensate, to a liner-axion-potential approach, familiar from string theory~\cite{silver}, which has been studied in \cite{Dorlis:2024yqw} using a dynamical systems approach. In the latter, inflation, characterised by the features proposed in \cite{bms,ms1,ms2}, to be described below, is guaranteed via an appropriate choice of boundary conditions for the pertinent algebraic system of differential equations. It must bs stressed, that the dynamical system analysis leads to a finite-duration inflation, and hence it excludes the Weyl ordering scheme, as mentioned above. 

Let us now review this scenario. The relevant gravitational effective action in (3+1)-dimensions, which describes the dynamics of the primordial Universe, according to the model of \cite{bms,ms1,ms2}, is given by a Chern-Simons (CS) gravity model~\cite{Jackiw,Alexander:2009tp}:
\begin{align}\label{sea}
S = \int d^4x \, \sqrt{-g}\, \Big[\frac{R}{2\kappa^2} - \frac{1}{2} \partial_\mu b \, \partial^\mu b - A \, b \, \mathcal R_{\rm CS}
\Big]\,,
\end{align}
where we work with the metric-signature convention $(-, +,+,+)$, and the curvature-tensor conventions of \cite{Dorlis:2024yqw}. 
The parameter
$\kappa = M_{\rm Pl}^{-1}$ is the gravitational constant in (3+1)-dimensional spacetime, with $M_{\rm Pl}=2.4 \times 10^{18}~\rm GeV$ the reduced Planck mass, $R$ is the Ricci scalar, $A$ is the coupling constant of mass dimension $-1$, and  
\begin{align}
    \mathcal R_{\rm CS} = \frac{1}{2} R^\mu_{\,\,\,\nu\rho\sigma} \, \widetilde R^{\nu \,\,\,\rho\sigma}_{\,\,\mu}\,,
\end{align}
is the CS gravitational anomaly, which is a total derivative. The field $b(x)$ is a pseudoscalar (axion-like) field. The quantity 
$\widetilde R_{\alpha\beta\gamma\delta} = \frac{1}{2} R_{\alpha\beta}^{\,\,\,\rho\sigma} \, \varepsilon_{\rho\sigma\gamma\delta}$ is the dual of the Riemann curvature tensor in (3+1)-dimensions, with $\varepsilon_{\rho\sigma\gamma\delta}= \sqrt{-g} \,\widehat \epsilon_{\alpha\beta\gamma\delta}$ the gravitationally covariant Levi-Civita density, with $\widehat \epsilon_{\alpha\beta\gamma\delta}$ the flat Minkowski spacetime Levi-Civita totally antisymmetric symbol, with the convention
$\widehat \epsilon_{0123}=+1,$ {\it etc.}

The action \eqref{sea} arises in the low-energy limit of bosonic string theory~\cite{str1,str2,kaloper}, in which case the coupling constant $A$ is given by~\cite{kaloper,bms} 
\begin{align}\label{Aval}
A= \sqrt{\frac{2}{3}}\, \frac{\alpha^\prime}{96\,\kappa}\,,
\end{align}where $\alpha^\prime = M_s^{-2}$ is the string Regge slope, with $M_s$ the string mass-scale parameter, which in general is different from the reduced Planck mass, and is treated here as a phenomenological parameter to be determined~\cite{Dorlis:2024yqw,Dorlis:2024uei} by comparing the predictions of the model, as far as inflation is concerned, to the data~\cite{Planck:2018vyg}. In the context of the cosmological model of \cite{bms,ms1,ms2} (which is called 
stringy RVM, in view of its aforementioned connection to the RVM), the action \eqref{sea} arises from the assumption that in the primordial, pre-inflationary Universe, only fields from the massless gravitational multiplet of the  underlying string theory ({\it i.e.} graviton, dilaton, and antisymmetric-tensor fields) are allowed to exist as external-line fields in the pertinent Feynman diagrams, describing the (perturbative) quantum gravity dynamics. In the case of superstrings, such fields are also the bosonic components of the superstring ground-state. In \cite{bms,ms1,ms2} it is assumed that 
supergravity is dynamically broken at the primordial Universe, so that the fermionic supersymemtric partners of the fields of the bosonic gravitational string multiplet acquire masses near the Planck mass, and thus decouple from the spectrum, leaving 
the effective action \eqref{sea} to describe the pre-inflationary dynamics of this string-inspired Universe (the dilaton $\Phi$ is assumed stabilised to a constant value, $\Phi=\Phi_0$, so that the string coupling $g_s=\exp(\Phi_0)$ is set to its phenomenologically relevant value~\cite{str1,str2}). The axion-like field
$b(x)$ in this context is the (3+1)-dimensional dual (after string compactification) of the antisymmetric-tensor field strength $\mathcal H_{\alpha\beta\gamma}$:
$\partial_\alpha b(x) \propto \varepsilon_{\alpha\beta\gamma\delta}\mathcal H^{\beta\gamma\delta}$. In this case, the axion-like field $b(x)$ is the so-called string-model-independent axion, characterised all string theory models~\cite{svrcek}. This field is distinct from other axions that arise as a result of compactification. Because it is related to the field strength of the spin-1 antisymmetric-tensor field (which is also called Kalb-Ramond (KR) field) of the gravitational string multiplet, we shall call it from now on the gravitational-axion or KR axion, for brevity. It will play a crucial r\^ole in our inflationary scenario~\cite{bms,ms1,ms2,Dorlis:2024yqw,Dorlis:2024uei}, which we now proceed to analyse.

The presence of chiral GW leads to the formation of a condensate of the gCS anomaly term $\langle \mathcal R_{\rm CS} \rangle$. The detailed computation of \cite{Dorlis:2024yqw} is based on the framework of weak quantum gravity for the pertinent quantised chiral-GW (tensor) perturbations of the metric~\cite{Lyth:2005jf,Dorlis:2024yqw,Dorlis:2024uei}. The corresponding chiral (Left(L), Right (R)) graviton Fourier modes are:
\begin{align}\label{chiralgrav}
  \widehat{\widetilde h}_{\rm L, \vec k}  &= u_{\rm L, \vec k}\, \widehat \alpha_{\vec k}^{-} + u^\star_{\rm R, -\vec k}\, \widehat b^+_{-\vec k}\,, \nonumber \\ 
  \widehat{\widetilde h}_{\rm R, \vec k}  &= u_{\rm R, \vec k}\, \widehat b_{\vec k}^{-} + u^\star_{\rm L, -\vec k}\, \widehat \alpha^+_{-\vec k}\,, 
\end{align}
where $\widehat{\dots}$ denotes quantum operators, $\widetilde{\dots}$ Fourier modes, and, in the notation of \cite{Dorlis:2024yqw}, 
$u_{\ell, \vec k} \equiv \widetilde v_{\vec k}/z_{\ell, \vec k}$, 
with $\ell = L,R$, 
$z_{\ell,\vec{k}}(\eta)= a(\eta) \sqrt{1 -l_{\ell} l_{\vec{k}} \, L_{CS}(\eta)}$, 
$\ell_{\vec{k}}=1$ and $\ell_{-\vec{k}}=-1$, $\ell_R \,\ell_{\vec{k}}= \ell_L \, \ell_{-\vec{k}}=1$, with $L_{CS}(\eta)=k  \xi$, $\xi=  \frac{4 A  b^{\prime}\kappa^2}{a(\eta)^2}$,
where $a(\eta)$ is the scale factor of the Universe as a function of the conformal time $\eta$, a derivative with respect to which is denoted by a prime.

\begin{figure}[ht!]
    \centering
\includegraphics[width=0.8\textwidth]{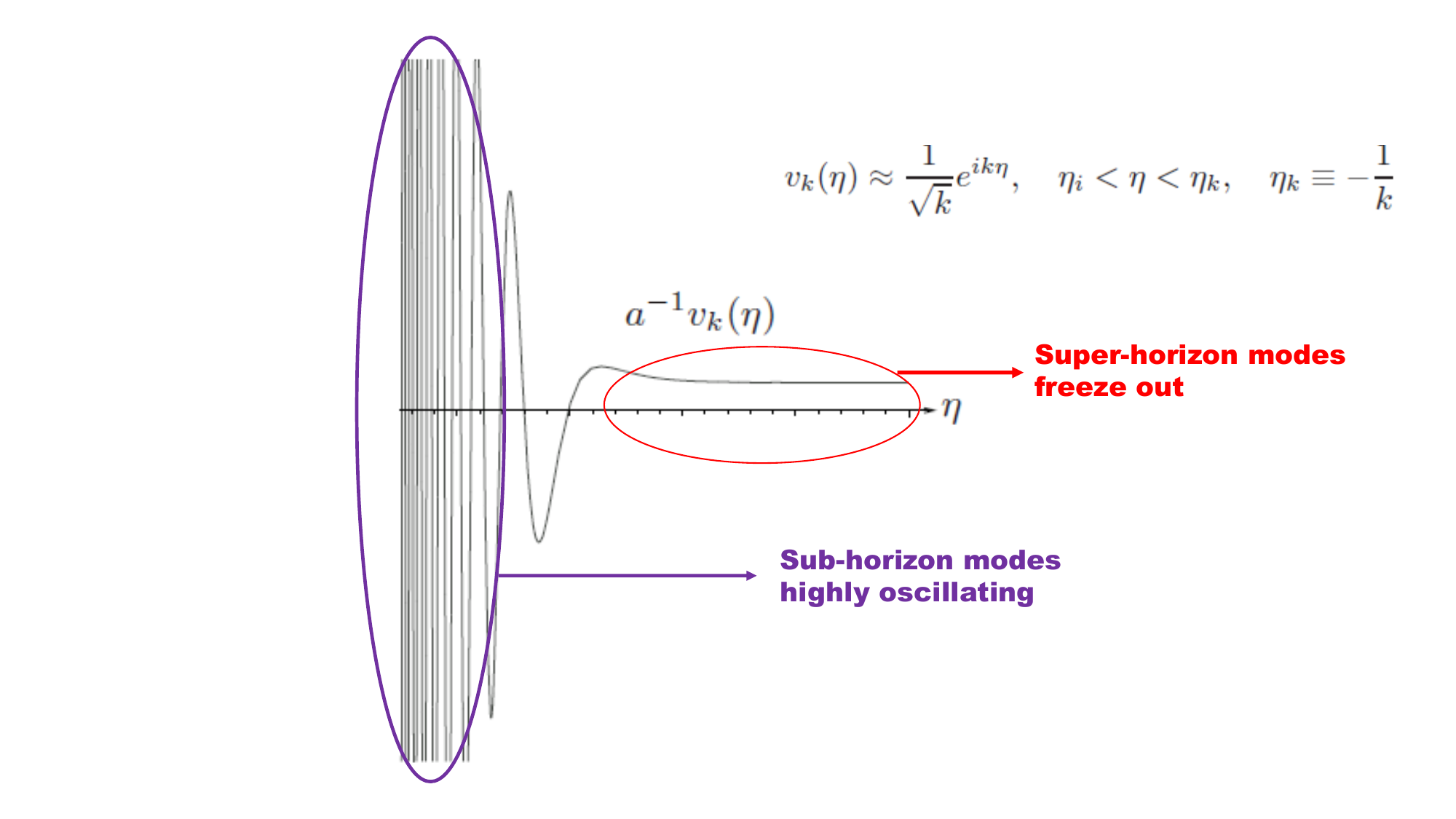}
    \caption{The subhorizon chiral gravitational wave modes oscillate highly, in contrast to the frozen superhorizon modes \cite{Mukhanov:2007zz}. In the above figure, $a$ denotes the scale factor of the Universe, $v_{\vec k}$ is the mode function, corresponding to a momentum scale $k$, and $\eta$ is the conformal time, with $\eta_i$ the value of the conformal time at the onset of inflation (end of stiff era in the model of \cite{bms,ms1}). Note here that the presence of the CS coupling will not lead to changes on the qualitative behavior for the sub and super horizon modes, but only to a small discrepancy between the amplitudes of the Left and Right modes (birefringence).} 
    \label{fig:horizon}
\end{figure}

The detailed computation of \cite{Dorlis:2024yqw}, using an ultraviolet (UV) cutoff $\mu$ for the momenta of the (quantum) graviton modes, 
yielded the following expression for the gCS condensate during the inflationary era, in which the Hubble parameter $H_I$ is almost constant:
\begin{align}\label{gCScond}
    \langle A\, \mathcal R_{\rm CS} \rangle^I_{\mathcal N_I} = -\mathcal{N}_I\frac{ A^2\,  \kappa ^4 \mu ^4}{\pi ^2}\dot{b}_I H_{I}^3\,,
\end{align}
where
\begin{equation}
    \dot{b}_I\sim 10^{-1} H_I M_{\rm Pl}\,,
    \label{bdotInflation}
\end{equation}
is the rate of change of the KR axion during the inflationary era. This rate can be justified using a dynamical-system approach to inflation in this model~\cite{Dorlis:2024yqw}, to be discussed below. 
The $\langle \dots \rangle $ denote $\langle 0|\dots |0\rangle $ with respect to the appropriate vacuum, which for both the stiff era and inflationary eras is taken to be the Bunch-Davies vacuum~\cite{Dorlis:2024yqw}, for which 
$\widehat{\alpha}^-_{\vec k} |0\rangle =0$. This is a vacuum of the sub-horizon modes, which are highly oscillating (see figure~\ref{fig:horizon} \cite{Mukhanov:2007zz}) and as such are the more energetic modes, in contrast to the superhorizon ones, which freeze out, during inflation, and re-enter after inflation. Due to this behavior, the dominant contribution to the anomaly condensate comes from the subhorizon modes, as can be seen from the following expression, 
\begin{equation}
     \langle R_{CS}\rangle = \frac{2}{\alpha^4}\int^{\alpha\mu}\frac{d^3 \Vec{k}}{(2\pi)^3}\left[ k^3  \left(u_{L,\vec{k}}  u^{*\prime}_{L,\vec{k}}-u_{R,\vec{k}}  u^{*\prime}_{R,\vec{k}} \right) + k \left(u^{\prime\prime}_{R,\vec{k}}u^{*\prime}_{R,\vec{k}}-u^{\prime\prime}_{L,\vec{k}}u^{*\prime}_{L,\vec{k}}\right)\right]\,.\\
     \label{Rcs_condensate}
\end{equation}

We now remark that the explicit presence of $\mu$ in the condensate \eqref{gCScond} finds a natural explanation in the context of embedding the CS gravity model \eqref{sea} to UV complete string theory models, of which the gravitational model \eqref{sea} is viewed as an {\it effective} local, low-energy, field theory, which is well-defined for energy and momenta below the cutoff scale 
\begin{align}\label{muMs}
\mu = M_s\,.
\end{align}
The validity of the sub-Planckian conjecture requires $M_s \lesssim M_{\rm Pl}$. The presence of $\mathcal N_I >1$ in \eqref{gCScond} guarantees the self-consistency of \eqref{muMs}~\cite{Mavromatos:2022xdo}.

In \eqref{gCScond}, the quantity $\mathcal N_I$ denotes the number density of sources of chiral GW at the onset of inflation. In the model of \cite{bms,ms1,ms2}, the inflationary era is preceded by a stiff-epoch, characterised with an equation of state $w=+1$, which is dominated by the massless KR-axion $b$-fields~\cite{ms1}. During the early phases of that era, there are no GW formed, so the CS term in \eqref{sea} vanishes. At the end of that era, there is a significant formation of chiral GW, which leads to inflation, in the way discussed in detail in \cite{Dorlis:2024yqw}. At that late-stiff epoch, the quantum chiral GW computation of the condensate, in the presence of $\mathcal N_S$ number density of sources, following the same methods as in the inflationary era discussed above, yields:
\begin{equation}\label{condstiff}
\langle A\, \mathcal R_{\rm CS}\rangle^{\rm stiff}_
      {\mathcal N_S}= -\mathcal N_S \, \frac{30\sqrt{6} A^2\, \kappa ^3 \mu ^4}{\pi ^2} H_{\rm stiff}(\eta)^4\,, \qquad H_
  {\rm stiff} (\eta)=\frac{\sqrt{\eta_0}}{2  \eta^{3/2}}\,,
\end{equation}
where $\langle \dots \rangle $ is computed with respect to the same Bunch-Davis vacuum as in the inflationary era~\cite{Dorlis:2024yqw}, discussed above, $\eta$ is the conformal time and $\eta_0 >0$ a constant, such that the late stiff era is defined as the one corresponding to $\eta/\eta_0 \gg 1$.

The (approximate) constancy of the condensate \eqref{gCScond} during the entirety of the inflationary era requires the matching of this value with \eqref{condstiff} at the end of the axion-dominated stiff era, which yields 
\begin{align}\label{NSNI}
\frac{\mathcal{N}_I}{\mathcal{N}_S}\sim 7\cdot 10^2 \left(\frac{H_i}{H_I}\right)^4\,,
\end{align}
where $H_i$ the value of the Hubble parameter at the end of the stiff/beginning of the inflationary era, that is at the era of the formation of the gCS anomaly condensate.

\section{Dynamical System Approach to  Inflation}\label{sec:dynsys}

The analysis of the passage from the stiff to inflation era has been performed in \cite{Dorlis:2024yqw} using dynamical system analysis. Inflation with the correct phenomenological features~\cite{Planck:2018vyg}, such as number of e-foldings in the range $N=$ 50 - 60, is guaranteed by an appropriate choice of the initial conditions of the equations that comprise the dynamical system under consideration, which also determines $H_i$, as we now proceed to review briefly.

\begin{figure}[ht!]
    \centering
\includegraphics[width=0.8\textwidth]{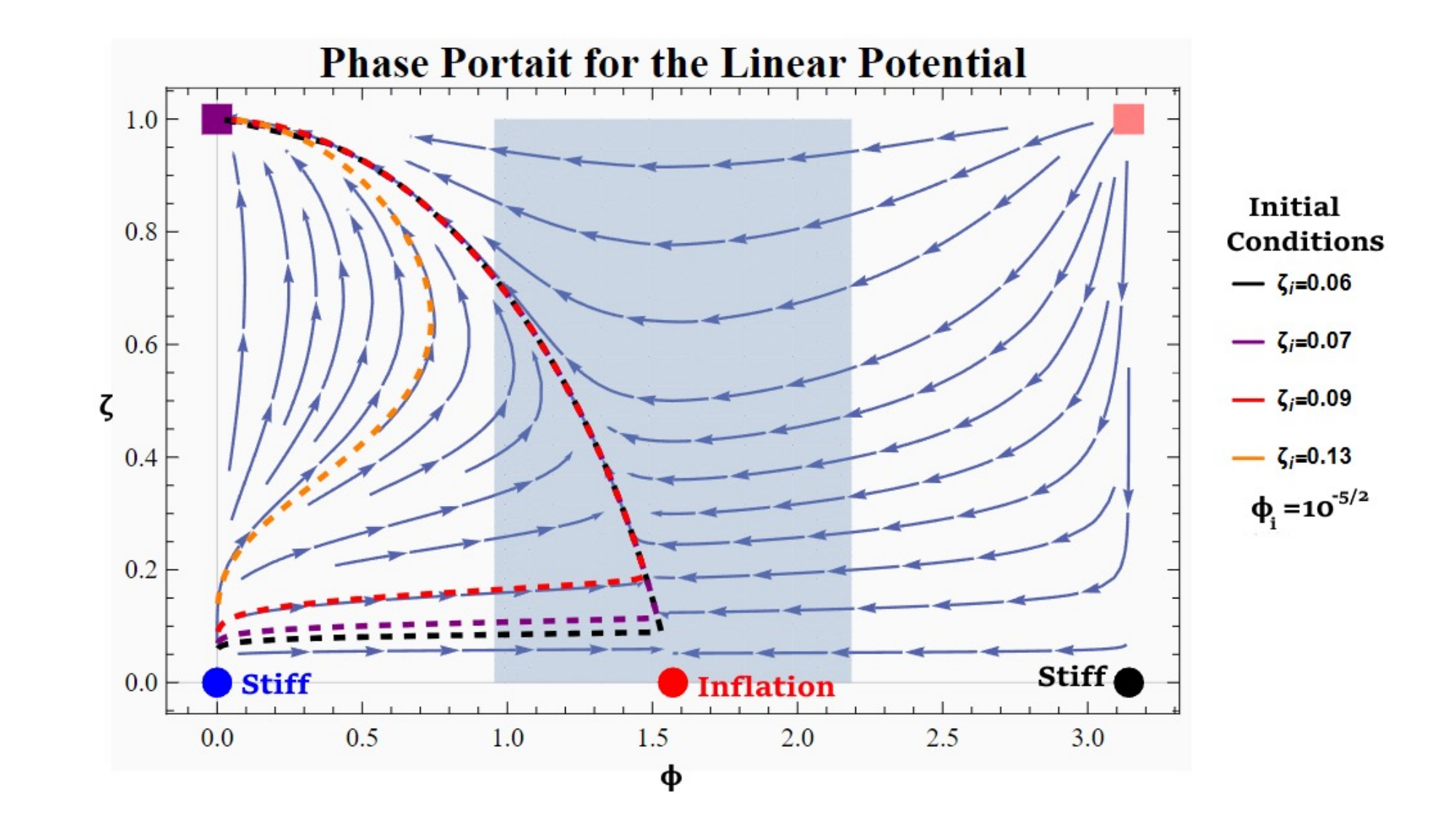}
    \caption{The phase portrait of the dynamical system \eqref{2ddynsys}, for the gCS anomaly condensate-induced potential \eqref{potb}. The critical points are marked on the Figure as coloured squares, or discs. The arrows indicate the direction of the flows towards or away from such points. The shaded rectangular region corresponds to accelerated expansion, with equation of state $\omega_b = x^2 - y^2 \leq-1/3$, with the (red coloured) disk in the center of the abscissa denoting exact exponential (de-Sitter type) expansion, $\omega_b=-1$. The various colour   lines correspond to orbits with different initial conditions, as indicated in the figure~\cite{Dorlis:2024yqw}.} 
    \label{fig:2dsys}
\end{figure}

\begin{figure}[ht!]
    \centering
\includegraphics[width=0.8\textwidth]{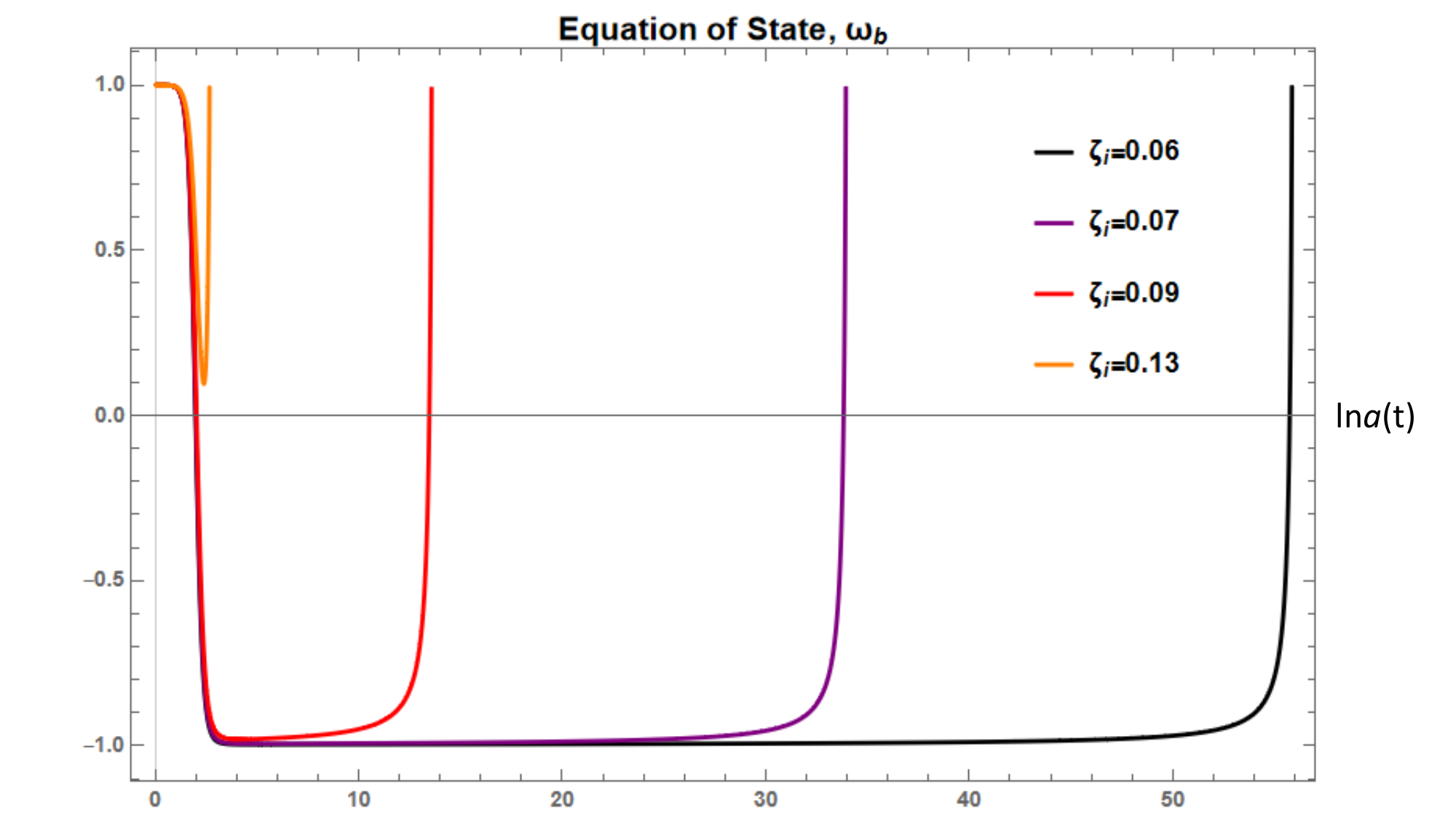}
    \caption{The Equation of state $w_b$ as a function of the e-foldings number $N=\ln a(t)$, for various initial conditions $\zeta_i$ for the parameter $\zeta$, and $\phi_i=10^{-5/2}$. The phenomenologically correct duration of inflation ($w_b \simeq -1$) in the range of $N =$ 50 - 60 e-foldings~\cite{Planck:2018vyg} occurs for the initial value $\zeta_i=0.06$~\cite{Dorlis:2024yqw}. Eternal inflation occurs for $\zeta_i \rightarrow 0^+$, that is,  $|b| \to \infty$.} 
    \label{fig:eos}
\end{figure}

\begin{figure}[ht!]
    \centering
\includegraphics[width=0.8\textwidth]{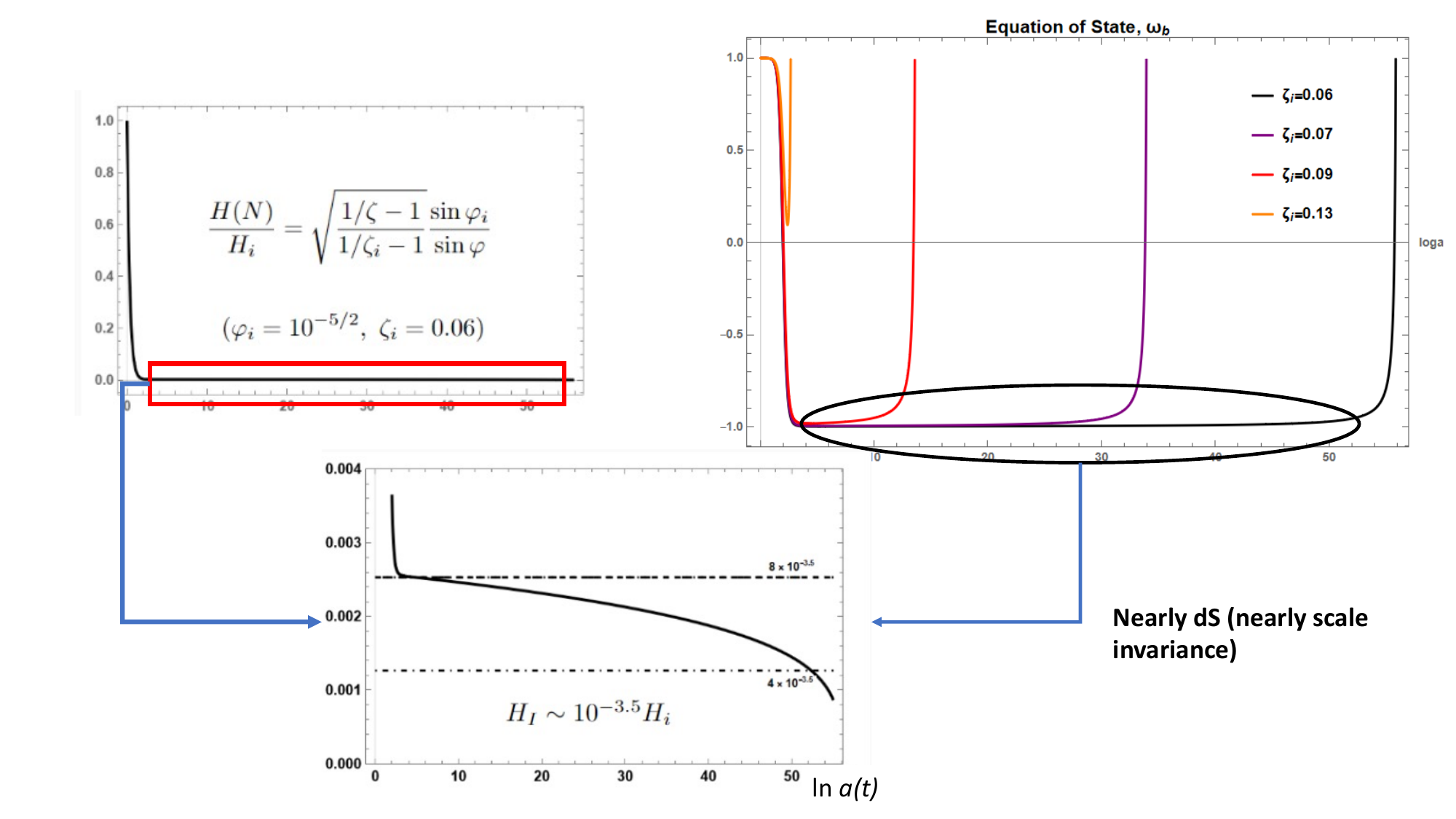}
    \caption{The evolution of the Hubble parameter during the short period from the end of the axion-dominated stiff era to the inflationary phase, in the stringy RVM cosmology of \cite{bms,ms1}.
    The Hubble rate drops by almost four orders of magnitude during that transition.}
    \label{fig:HEvol}
\end{figure}

The relevant algebraic system of equations follows from Einstein's equations for the action \eqref{sea}, under the assumption of a constant gCS condensate during inflation. Such an assumption leads to a linear axion-$b(x)$ potential in the effective action \eqref{sea}:
\begin{align}\label{potb}
   V(b) = b(x) \, A\langle \, \mathcal R_{\rm CS}\rangle^{I}_{\mathcal N_I} \equiv b(x) \, \mathcal C \,,
\end{align}
with $\mathcal C <0$ a constant given by \eqref{gCScond} (we treat here $H_I$ as approximately constant, during the entire duration of inflation. In actual terms it may vary slowly over the cosmic time in that phase of the Universe. The dynamics is described by the Lagrange equations for the graviton and axion fields stemming from \eqref{sea} in the isotropic and homogeneous Universe case~\cite{Dorlis:2024yqw}:
\begin{align}\label{eqssea}
 &  3H^2 = \kappa^2 \Big(\frac{\dot b^2}{2} + V(b)\Big)\,,  \nonumber \\
 & 2\dot H + 3H^2 = - \kappa^2 \Big(\frac{\dot b^2}{2} - V(b)\Big)\,,  \nonumber \\
&  \ddot b + 3 H\, \dot b + V(b)_{, b}
 = 0\,, \qquad V_{, b} \equiv \frac{\delta V(b)}{\delta b}\,,
\end{align}
where $V(b)$ is given in  \eqref{potb}, $H(t)$ is the Hubble rate, 
and the overdot denotes derivation with respect to the cosmic (Robertson-Walker frame) time $t$. The first of the above equation is nothing than the Friedman equation stemming from the temporal $00$ component of the Einstein's equations. The second correspond to the spatial-component of the Einstein equations, while the third represents the equation of motion of the axion $b(t)$ field. The required dynamical system follows from \eqref{eqssea} upon using the change of variables $x \equiv \kappa \, \dot b/(\sqrt{6}\, H)$, $y = \kappa \, \sqrt{|V(b)|}/(\sqrt{3}\, H)$, $\lambda = - V_{, b}/(\kappa \, V)$ and $\Gamma \equiv V\, V_{,bb}/V_{, b}^2$, which yields~\cite{Dorlis:2024yqw}:
\begin{align}\label{dynsys}
x^\prime &= -\frac{3}{2} \Big[ 2x - x^3 + x\, (y^2-1) + \sqrt{\frac{2}{3}}\, \lambda\, y^2\Big]\,, \nonumber \\
y^\prime &= -\frac{3}{2} \, y \, \Big[ -x^2 + y^2 -1 + \sqrt{\frac{2}{3}}\, \lambda\, x\Big]\,, \nonumber \\
\lambda^\prime &= -\sqrt{6}\, \Big(\Gamma -1 \Big)\, \lambda^2\, x\,,
\end{align}
with the prime in this section denoting differentiation with respect to the e-folding number $N = \ln a(t)$, where $a(t)$ is the scale factor of the Universe. Taking into account that in these variables the Friedman equation provides a constraint $x^2 + y^2 = 1$, the system of equations reduces to a two-dimensional one, which, upon using the parametrization $x=\cos \phi$, $y \equiv \sin \phi$ 
and changing variable $\zeta = \lambda/(\lambda + 1)$, reads:
\begin{align}\label{2ddynsys}
\phi^\prime &= \Big(3\, \cos\phi - \sqrt{\frac{3}{2}}\, \frac{\zeta}{\zeta-1} \Big)\, \sin\phi \,, \nonumber \\
\zeta^\prime & = -\sqrt{6}\, \Big(\Gamma -1\Big)\, \zeta^2\, \cos\phi
\,.
\end{align}
The phase portrait (in the $\zeta-\phi$ plane) corresponding to the two-dimensional system \eqref{2ddynsys}, with the condensate-induced potential \eqref{potb}, is given in figure \ref{fig:2dsys}, for $\phi_i = 10^{-5/2}$ and various boundary conditions $\zeta_i$ for the parameter $\zeta$, where the suffix $i$ denotes initial value. Here the word ``initial'' is interpreted as referring to the onset of inflation. In the model of \cite{bms,ms1} the initial value coincides with the end of the axion-dominated stiff era, which is succeeded by inflation, as mentioned above. 
As becomes obvious from the figure, the inflation critical point (red disc) is a {\it saddle} point of the flow. The equation of state as function of the e-foldings, for various boundary conditions is given in figure~\ref{fig:eos}. We observe that the 
duration of the inflationary phase (for which $w_b \simeq -1$ (approximately de Sitter spacetime)) increases with decreasing values of $\zeta_i$. For the boundary condition $\phi_i=10^{-5/2}$, the 
phenomenologically acceptable range for the duration of inflation, of order $N=$ 50 - 60 e-foldings~\cite{Planck:2018vyg} occurs for the initial value $\zeta_i = 0.06$~\cite{Dorlis:2024yqw}. 
Eternal de-Sitter inflation occurs for $\zeta_i \rightarrow 0^+$, which amounts to $|b| \to \infty$.
During the short period interpolating between the end of the stiff era to the inflationary phase, the Hubble rate is reduced significantly, from a value $H_i$ to the approximately constant de-Sitter value $H_I$ during inflation, according to the law
\begin{equation}
 \frac{H(N)}{H_i}= \sqrt{\frac{1/\zeta-1}{1/\zeta_i-1}}\frac{\sin\phi_i}{\sin\phi}\,,
\end{equation}
where the subscript $i$ denotes initial values. For the specific initial conditions leading to inflation:
$\phi_i=10^{-5/2},\, \,\zeta_i=0.06$,
the evolution of the Hubble parameter $(t)$ is depicted in figure~\ref{fig:HEvol}. 

As we observe, $H$ decreases very fast, until it acquires an approximately constant value, signifying the entrance of the Universe to an approximately de Sitter, inflationary, phase, $H_I\approx {\rm constant}$, for $N>50$ e-folds, with an order of magnitude, 
\begin{align}\label{HIi}
H_I\sim 10^{-3.5}H_i\,.   
\end{align}
To get agreement with cosmological data~\cite{Planck:2018vyg}, one should 
set as an upperbound 
\begin{align}\label{HI}
H_I \lesssim 10^{-5}\, M_{\rm Pl}\,.
\end{align}
From the result \eqref{HIi}, we have, on account of \eqref{NSNI}, that to guarantee anomaly-condensate inflation, respecting the sub-Planckian conjecture, 
we need to increase the number of chiral GW sources at the end of the stiff era, as:
\begin{align}\label{sourcesNINS}
    \frac{\mathcal{N}_I}{\mathcal{N}_S}\sim 7\cdot 10^{16}\,.
\end{align}
Using the expressions~\cite{Dorlis:2024yqw}
$\frac{b}{M_{\rm Pl}}=\frac{\zeta - 1}{\zeta}$ and $\frac{\dot{b}}{H M_{\rm Pl}}= \sqrt{6} \cos\phi$, which follow from the previously mentioned definitions, one can estimate that, for the boundary conditions ($i$) leading to inflation, the following order of magnitude estimates occur:
\begin{equation}
    \frac{{|b_i|}}{M_{\rm Pl}} \sim \mathcal{O}\left(10\right)\,, \quad  \frac{\dot{b}}{H_I M_{\rm Pl}} \sim \mathcal{O}\left(10^{-1}\right)\,,
    \label{b_bdot_orders}
\end{equation}
which remain approximately constant during the entirety of inflation. Eq.~\eqref{b_bdot_orders} yields \eqref{bdotInflation}, mentioned previously. Such values had been conjectured in \cite{bms,ms1}, in order to qualitatively ensure the approximate constancy (during inflation) of the effective cosmological constant in the action \eqref{sea} under the formation of the condensate $\Lambda_{\rm eff} = A\, b \langle \mathcal R_{\rm CS} \rangle \simeq $ constant. Above, these values have been derived within the dynamical system approach to the linear-axion-potential induced inflation. 

\section{Chiral-Gravitational-Wave Sources}\label{sec:gworigins}

The nature of the possible sources of chiral GW in the early universe constitutes an important question, especially if we aim to enhance their production during inflation and satisfy \eqref{sourcesNINS}. There is a variety of such sources, ranging, for instance, from populations of (rotating) primordial black holes, in the vicinity of which the gCS anomaly term is non-zero~\cite{kaloper,Yunes:2009hc,Yagi:2012ya,Chatzifotis:2022mob}, to non-spherically-symmetric collapsing or colliding domain walls in non-equilibrium-phase-transition situations with a biased double-well potential, where the bias might be even statistical in origin, due to unequal occupation numbers of the two vacua~\cite{Lalak:1994qt,Coulson:1995nv}. Such scenarios may be in operation during the pre-inflationary phase of the stringy RVM!\cite{ms1}. 

Below we shall discuss briefly another kind of sources of chiral GW, 
that of  axionic domain walls, which would also fit 
our stringy RVM, upon some conditions that we shall discuss below. This mechanism does not involve biased potential wells. The behavior of GW passing through axionic domain walls in the context of CS gravity has been studied recently in \cite{KannoSoda}. In the context of our string-inspired cosmology,the axionic domain wall background might be associated with compactification axions that are abundant in string theory~\cite{svrcek}, and are distinct from the KR axion $b$ field (the latter does not have a potential,
in the absence of chiral GW, which is crucial for the pre-inflationary stiff phase~\cite{ms1}).

Assuming a dominant compactification-axion species, $a(x)$, the dynamics of the axionic domain wall is described by the action
\begin{equation}
\label{action_axion}
    S_{DW}[a]=-\int d^4 x \sqrt{-g}\left(\frac{1}{2}\partial_\mu a \ \partial^\mu a  \ + \ V(a)\right) \ , 
\end{equation}
with $V(a)$ denoting the double well potential 
\begin{equation}
\label{double_well_potential}
    V(a)=\frac{\lambda}{4}\left(a^2 - v^2\right)^2 \ ,
\end{equation}
which has two degenerate minima, located at $a=\pm v$, and $\lambda$ denotes a coupling constant. Such a potential requires some discussion in view of a possible embedding of this model for GW sources into string theory.

Indeed, in view of the swampland criteria for a consistent formulation of an effective quantum-gravity (QG)~\cite{swamp1,swamp2,swamp3,swamp4,swamp5}, a cosmological constant, as the one appearing naively in the potential  \eqref{double_well_potential}, is not allowed, unless the local minimum  corresponding to it is {\it metastable}, that is it has a finite duration, and eventually the system goes out of this phase. In our point of view, should the potential \eqref{double_well_potential} characterise effective field theories from strings, as required for an embedding of the model into the stringy RVM~\cite{ms1}, there must be non-trivial imaginary parts in the corresponding low-energy effective action, in a similar way with our condensate inflation~\cite{Dorlis:2024uei}. Only in this sense, this kind of axionic domain-wall mechanism for chiral GW  production will be compatible with our string-effective theory considerations. In our phenomenological discussion below, we assume this to be the case.

Upon assuming a planar domain wall to be in the
$(x-y)$-plane and being orthogonal to the $z$-axis, the equations of motion stemming from the action \eqref{action_axion} are given by:
\begin{equation}
    \label{e.o.m. axion}
    a^{\prime\prime}(z) = \lambda \ a(z)   \left(a(z)^2 - v^2 \right) \,,
\end{equation}
where a prime in this section denotes differentiation with respect to the $z$ coordinate, $a^\prime \equiv \partial_z a$.
Eq.~\eqref{e.o.m. axion}  admits the following domain wall solution:
\begin{equation}
    \label{solution_domain_wall}
    a(z)=v \tanh\left(\frac{\sqrt{\lambda}}{2}  v \,  z\right) \ ,
\end{equation}
and the surface energy density given by
\begin{equation}
    \sigma=\int^{\infty}_{0} d z \ (a^{\prime}(z))^2 \sim \sqrt{\lambda} \ v^3 \,.
\end{equation}
The axionic domain wall background configuration of \eqref{solution_domain_wall} represents a local structure and that's why we shall  proceed and consider GW in the Minkowski spacetime, expressed by the tensor perturbation of the background metric:
\begin{equation}
\label{metric+flat_perturbation}
    ds^2 = -dt^2 + (\delta_{ij} + h_{ij})dx^i dx^j \ ,
\end{equation}
where $\delta_{ij}$ is the Kronecker delta. The action is given by \eqref{sea}, with the main difference being that now we are dealing with a local axionic domain wall  background given by \eqref{solution_domain_wall} and the perturbation of the gCS term is performed around a flat background \eqref{metric+flat_perturbation}. The GW passing through the axionic domain wall are described by the following action \cite{KannoSoda}:
\begin{align}
    S_{GW} &= \frac{1}{8\kappa^2}\int d^4x \;\Big[ \left(\dot{h}_{ij} \dot{h}^{ij} -\partial_k h_{ij}\partial^k h^{ij}\right) \nonumber \\ & ~+ 4\,\widetilde{A}\, \kappa^2\, a^\prime (z) \   \hat{\varepsilon}^{zjk}\left(\partial_l \dot{h}^{ m}_{ \ k}\partial^l h_{jm}-\partial_l h^m_{ \ j}\partial_m \dot{h}^{l}_{ \ k} - \dot{h}_{jl}\ddot{h}^{l}_{ \ k}\right)   \Big] \ ,
    \label{ActionTensor}
\end{align}
where $\widetilde{A}$ denotes the coupling of the (compactification) axion field $a(x)$ to the gCS anomaly $\mathcal R_{\rm CS}$, with $\widetilde A \ne A$, and 
$\hat{\varepsilon}^{ijk}$ denotes the three dimensional Levi-Civita symbol and $a=a(z)$. In the helicity basis, the equations of motion stemming from \eqref{ActionTensor} for Left and Right handed modes are decoupled (something that does not happen in the $(+,\times)$ basis), and are given by:
\begin{equation}
\label{e.o.m._for_HL_and_HR}
    \left(1 \ \pm \ 4 \, \widetilde{A}\, \kappa^2\omega \ a^{\prime}\right)\mathcal{H}^{\prime \prime}_{L/R} \ \pm \ \left(4 \, \widetilde{A}\,\kappa^2\omega \ a^{\prime \prime}\right) \mathcal{H}^{\prime}_{L/R} \ + \ \omega^2\left(1 \ \pm \ 4\, \widetilde{A}\,\kappa^2\omega \ a^{\prime }\right)\mathcal{H}_{L/R}=0 \ ,
\end{equation}
where $h_{L/R}=\mathcal{H}_{L/R}(z,\omega) \ e^{i \omega t}$. The above equation \eqref{e.o.m._for_HL_and_HR} demonstrates that, the background axionic domain wall, serves (because of the interaction with the CS-gravitational term) as a source of chiral gravitational waves, without the need for any bias in the potential, though. Performing a simple field redefinition in the following way:
\begin{equation}
\label{redefinition_functions}
    \mathcal{H}_{L/R}=\frac{\psi_{L/R}}{\sqrt{F_{L/R}}} \ \ , \text{where} \ \ F_{L/R}=1 \ \pm \ 4\, \widetilde{A}\, \kappa^2\omega \ a^{\prime}(z) \ ,
\end{equation}
the equations of motion \eqref{e.o.m._for_HL_and_HR} become:
\begin{equation}
\label{e.o.m.for_Psi}
    \psi_{L/R}^{\prime \prime} +\left[\omega^2 - V^{eff}_{L/R}(z)\right]\psi_{L/R} =0 \ ,
\end{equation}
where the effective potential $V^{eff}_{L/R}$ is given by:
\begin{equation}
\label{effective_potential}
    V^{eff}_{L/R}(z,\omega)=-\frac{1}{4}\left(\frac{F_{L/R}^\prime}{F_{L/R}}\right)^2 + \frac{1}{2}\frac{F_{L/R}^{\prime\prime}}{F_{L/R}} \ .
\end{equation}
\begin{figure}[ht]
    \centering
\includegraphics[width=\textwidth]{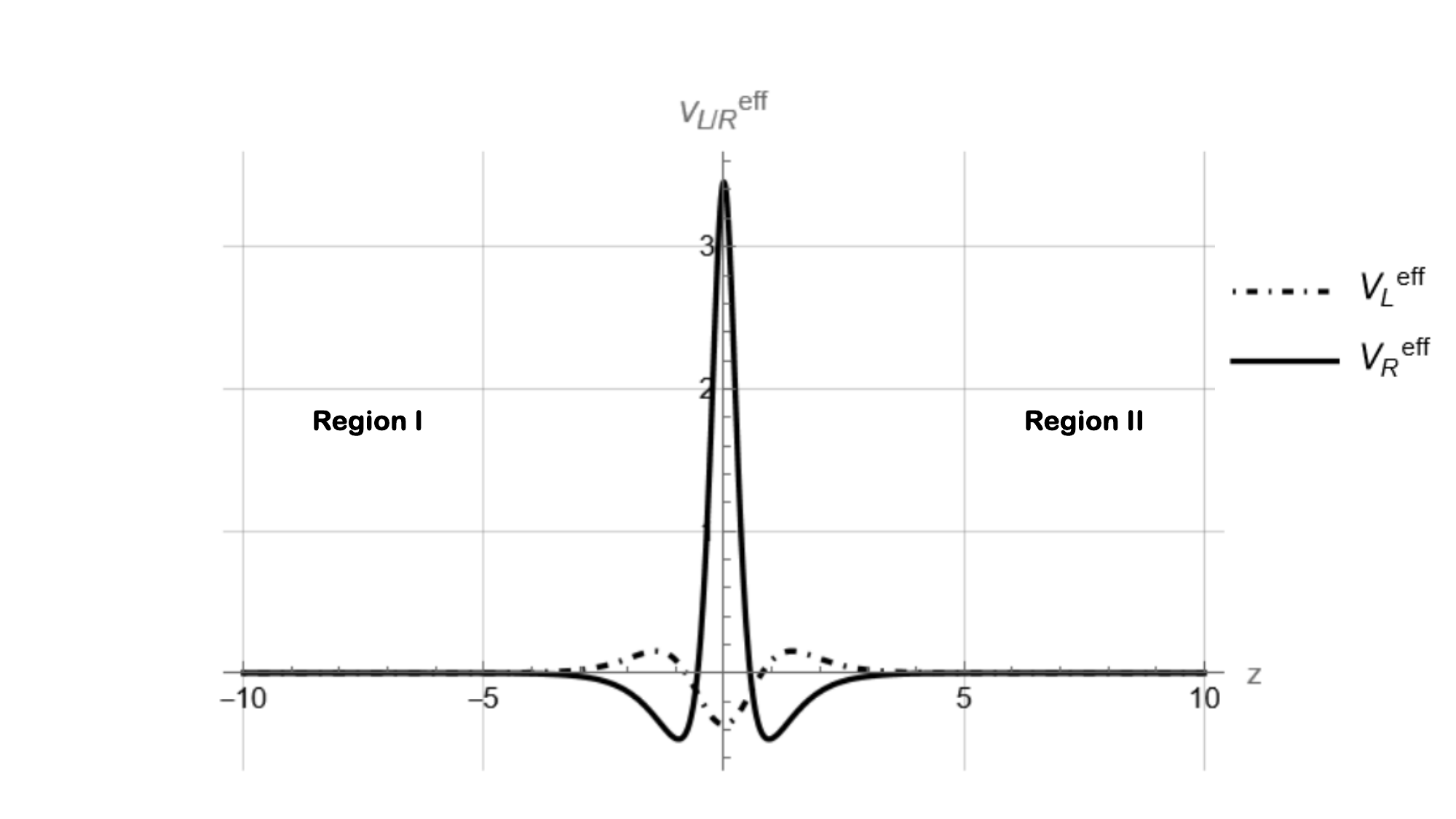}
    \caption{The effective potential for left and right handed modes. The values chosen for the plot are $ v=0.9, \ \lambda=2$ and $4A\kappa^2 \omega =1$  \cite{KannoSoda}.}
\label{Effective_potential_figure}
\end{figure}
The effective potential is plotted in figure \ref{Effective_potential_figure}, where we can see that in the far left and far right zones (denoted as Regions I and II, respectively in the figure), the potential vanishes. In these limits, the redefinition functions $F_{L/R}\to 1$ and there is no distinction between left-handed and right-handed modes, in other words, far from the potential, the effects of chirality are vanishing. The different behavior for Left and Right modes stemming from equation \eqref{e.o.m.for_Psi} results in different transmission coefficients through the potential barrier, different for each polarization. Consequently, such a behavior will yield a non - vanishing value for the condensate $\langle \mathcal{R}_{CS}\rangle \neq 0$, which will depend on the details of the source, that is, the axionic domain wall in this case.
\color{black}

\section{Instabilities of the Inflationary Vacuum}\label{sec:instab}

From a weak quantum-gravity field-theory point of view, adopted in \cite{Dorlis:2024yqw,Dorlis:2024uei}, the computation of the gCS condensates \eqref{gCScond}, \eqref{condstiff}, which are composite graviton operators,   
implies the presence of imaginary parts in all quantum-operator ordering schemes, except the Weyl (symmetric) one. Specifically, the analysis of \cite{Dorlis:2024uei} yields for the second-quantised gCS terms:
{\small \begin{align}
     \widehat{R_{\mu\nu\rho\sigma}^{\ \ \ \ \ \ \ \ \ \star} R^{\nu\mu\rho\sigma}} - 
      \widehat{( R_{\mu\nu\rho\sigma}^{\ \ \ \ \ \ \ \ \ \star}R^{\nu\mu\rho\sigma})}^\dag  &= 4 \int \frac{d^{3}\vec{k}d^{3}\vec{k'}}{(2\pi)^3} \ e^{i(\vec{k}+\vec{k'}) \cdot \vec{x}}\left\{ k^2 k' \ell_{\vec{k'}}\left( \left[\hat{h}_{L,\vec{k}},\hat{h}^{\prime}_{R,\vec{k'}}\right] + \left[\hat{h}^{\prime}_{L,\vec{k'}},\hat{h}_{R,\vec{k}}\right] \right) \right\} \nonumber \\ 
      &- 
       4 \int \frac{d^{3}\vec{k}d^{3}\vec{k'}}{(2\pi)^3} \ e^{i(\vec{k}+\vec{k'}) \cdot \vec{x}}\left\{ k' \ell_{\vec{k'}} \left( \left[\hat{h}^{\prime \prime}_{L,\vec{k}},\hat{h}^{\prime}_{R,\vec{k'}}\right] + \left[\hat{h}^{\prime }_{L,\vec{k'}},\hat{h}^{\prime\prime}_{R,\vec{k}}\right] \right)  \right\} \,,
       \label{nonhermitian_Rcs}
\end{align}}
where the $\dagger$ is the Hermitian conjugate, the $\widehat{\dots}$ denotes quantum operators, and the $\star $ is the Hodge-star operation, that is the dual in flat Minkowski spacetime. The important point to notice is that the imaginary paerts are related exclusively to quantum operators and thus are due to quantum physics effect, which destabilize the condensate-induced inflationary vacuum. In view of the interpretation of such imaginary parts, as implying instabilities of the respective ground state, and thus a finite life time for inflation, 
these ordering schemes
would correspond to the aforementioned dynamical system cases, which imply finite life times for the respective inflationary scenarios, with the Weyl scheme corresponding to the eternal de Sitter case, corresponding to $\zeta \to 0^+$ case.

In a particular ordering scheme, specified in \cite{Dorlis:2024uei}, the imaginary parts of the gCS condensate during inflation have been estimated to leading order in small quantities, as:
\begin{align}\label{imRCS}
 2 {\rm Im}\langle \widehat{\mathcal R_{\rm CS}}\rangle = {\rm Im}\langle \widehat{R_{\mu\nu\rho\sigma}\, \widetilde R^{\nu\mu\rho\sigma}} \rangle = 
 \frac{16  A \text{$\dot{b}$} \mu ^7}{7 M^{4}_{\rm Pl} \ \pi ^2} \left[1+\left(\frac{H_I}{\mu}\right)^2\left(\frac{21}{10}-6\left(\frac{A\mu\text{$\dot{b}$}}{M_{\rm Pl}^2}\right)^2\right)\right] \ , 
\end{align}
where $\mu = M_s$ ({\it cf.} \eqref{muMs}). Hence, in view of \eqref{HI}, and $\eqref{muMs}$,
the second term inside the square brackets 
on the right-hand side of the above equation is subdominant, and will not play a r\^ole in our estimates of the inflationary vacuum lifetime. The latter is estimated by observing~\cite{Dorlis:2024uei} that the imaginary parts \eqref{imRCS} induce imaginary parts in the Hamiltonian $\mathcal H$ of GW from a single source, 
\begin{equation}\label{imH}
 {\rm Im}\left(\mathcal H\right) = \int d^3 x \ \sqrt{-g}\,  \frac{1}{2} A \ b \ {\rm Im} \left(\langle R_{\mu\nu\rho\sigma}\widetilde{R}^{\nu\mu\rho\sigma}\rangle \right) \approx V^{(3)}_{dS} \ \frac{8  b A^2  \text{$\dot{b}$}  \mu ^7}{7 M^{4}_{\rm Pl} \ \pi ^2}~
\end{equation}
where $\approx$ implies that we keep only leading order contributions in the small quantity $\kappa^2 \dot b$, and $V^{(3)}_{dS}$ denotes the de Sitter 3-volume of the inflationary spacetime, 
\begin{equation}
   V^{(3)}_{dS} T^E =  \frac{24 \pi^2}{M^{2}_{\rm Pl} \Lambda} \ , \quad \Lambda =  3 H^{2}_I~,
   \label{4volume}
\end{equation}
with $T^E$ corresponding to the Euclidean time defining the appropriate duration of inflation, which phenomenologically is of order~\cite{Planck:2018vyg} 
\begin{align}\label{durinfl}
T^E\sim (50-60)H_I^{-1}\,,
\end{align}
The resulting life time  
of the inflationary vacuum is then given by the inverse of \eqref{imH}, in units of $\hbar=1$, $\tau = ({\rm Im}(\mathcal H))^{-1}$. On requiring $\tau = T^E $ to be of the phenomenologically expected order \eqref{durinfl}, 
one then obtains~\cite{Dorlis:2024uei}
\begin{equation}
\frac{M_s}{M_{\rm Pl}}\lesssim
0.215\,,
\end{equation}
which is quite consistent with the findings of the linear-axion-potential dynamical-system analysis of \cite{Dorlis:2024yqw}, 
described above, which concentrated on the real part of the CS condensate. 

\section{Slow-roll Inflationary Phenomenology}\label{sec:slowroll} 

The slow-roll phenomenology of the axion-driven inflation has been discussed in \cite{Dorlis:2024uei}, and we concluded that the spectral index $n_s$ can become much closer to the central value, indicated by observations~\cite{Planck:2018vyg}, if one includes periodic  modulations of the condensate-induced linear-axion-$b$ potential \eqref{potb}, which, in the stringy RVM cosmology of \cite{bms,ms1} might be induced at the end of the stiff-era via non-perturbative instanton configurations of an appropriate grand unified gauge group that characterises the underlying microscopic string theory model:
\begin{align}\label{axionperiodpot}
&V_{\rm eff}(b) = b(x) \, \Lambda_{\rm cond}^3 + 
V_{\rm eff}^{\rm periodic} (b)
\,, \qquad \Lambda_{\rm cond}^3  \equiv  A\, \langle \mathcal R_{\rm CS}\rangle^{I}_{\mathcal N_I}\,, \nonumber \\
&V_{\rm eff}^{\rm periodic} (b) = \Lambda_1^4 \, {\rm cos}\Big(16\pi^2 A \, b(x)\Big) \equiv \Lambda_1^4 \, {\rm cos}\Big(\frac{b(x)}{f_b}\Big)\,,
\end{align}
where $\Lambda_1$ is the scale of the gauge-group instantons, and $f_b $ is the axion coupling. 
In the StRVM, $f_b$ is given by \eqref{Aval}~\cite{ms1,kaloper,Dorlis:2024uei}:
\begin{align}\label{fbval}
f_b = 0.37\, \frac{M_s^2}{M_{\rm Pl}}\,.
\end{align}
The scale $\Lambda_1$ depends on the particular gauge model considered. For our purposes here it 
is treated as a phenomenological parameter of our low-energy string-inspired effective theory. The  scale $\Lambda_1$  is suppressed by the exponential of the (Euclidean) one-instanton action, $\mathcal S_{\rm inst}$, 
\begin{align}\label{L1}
\Lambda_1 = \xi \, M_s \, \exp\Big(-\mathcal S_{\rm inst}\Big) \lesssim \xi \, M_s \, \exp\Big(-\frac{8\pi^2}{g^2_{\rm YM}}\, |n|\Big) \ll M_s\,,
\end{align}
where $\xi >0$ is a numerical factor, that can in principle be calculated within a specific gauge theory. For our purposes we assume $\xi =\mathcal O(1)$. The instanton action $S_{\rm inst}$ is known to be bounded from below by~\cite{Coleman:1978ae,Diakonov:2002fq,Tong:2005un}):
\begin{align}\label{instbound}
\mathcal S_{\rm inst} \gtrsim \frac{8\pi^2}{g^2_{\rm YM}} |n| \,,
\end{align}
where $g_{\rm YM}$ is the Non-Abelian gauge group (Yang-Mills) coupling, and,  $n$ labels the respective topological instanton sector.

For the periodically-modulated axion potential \eqref{axionperiodpot}
the corresponding inflationary slow roll parameters read~\cite{Dorlis:2024uei}:
\begin{align}\label{eiperiodic}
\epsilon_1 & \equiv \frac{M_{\rm Pl}^2}{2} \, \Big(\frac{V^\prime}{V} \Big)^2 \simeq \frac{M_{\rm Pl}^2}{2}\, \Big[\frac{\Lambda_{\rm cond}^3 - \frac{\Lambda_1^4}{f_b}\, {\rm sin}\Big(\frac{b}{f_b}\Big)}{b(0)\, \Lambda_{\rm cond}^3 + \Lambda_1^4\, {\rm cos}\Big(\frac{b(0)}{f_b}\Big)}\Big]^2 \nonumber \\&\simeq \frac{M_{\rm Pl}^2}{2\, b^2(0)} \, \Big[ 1 + \frac{2\, \Lambda_1^4\, {\rm sin}\Big(\frac{b(0)}{f_b}\Big)}{f_b|\Lambda_{\rm cond}^3|} - \frac{2\, \Lambda_1^4\, {\rm cos}\Big(\frac{b(0)}{f_b}\Big)}{|b(0)\,\Lambda_{\rm cond}^3|} + \dots \Big] \nonumber \simeq 
\frac{M_{\rm Pl}^2}{2\, b^2(0)}\,,  
\nonumber \\
\epsilon_2 & \equiv 
2M_{\rm Pl}^2 \,  \Big(\frac{V^\prime}{V} \Big)^2 - 2 M_{\rm Pl}^2\, \frac{V^{\prime\prime}}{V} \equiv 4 \epsilon_1 - 2 \eta
\nonumber \\
&\simeq  2\frac{M_{\rm Pl}^2}{b^2(0)} + 2 M_{\rm Pl}^2 \, \frac{\Lambda_1^4}{f_b^2}\, \frac{{\rm cos}\Big(\frac{b(0)}{f_b}\Big)}{|b(0)\, \Lambda_{\rm cond}^3|} \simeq  2\frac{M_{\rm Pl}^2}{b^2(0)} + 2 \frac{M_{\rm Pl}^2}{f_b^2}\, \frac{\Lambda_1^4}{|b(0)\, \Lambda_{\rm cond}^3|}\,,
 \end{align}
where we took into account that 
$\Lambda_1 \ll \Lambda_{\rm cond}, \, f_b\,, M_{\rm Pl}, M_s$, as a result of the exponential suppression of the scale $\Lambda_1$  by the large instanton action \eqref{instbound}, and in the last equality of the second line we have made the order of magnitude assumption that ${\rm cos}\Big(\frac{b(0)}{f_b}\Big) = \mathcal O(1)$.

The spectral index is given by~\cite{Planck:2018vyg}
\begin{align}\label{scalarspectral}
n_s \simeq 1 - 2\epsilon_1 - \epsilon_2\,.
\end{align}
The Planck-collaboration constraints on inflation~\cite{Planck:2018vyg}, after taking into account lensing and Baryon-acoustic-oscillation (BAO) measurements, indicate that: 
\begin{align}\label{nsplanck}
n_s & = 0.9649 \pm 0.0042 \quad (\rm{68\%~C.L.,~Planck~TT,~TE,~EE~+~lowE~+~lensing})\,, \nonumber \\
n_s & = 0.9665 \pm 0.0038 \quad (\rm{68\%~C.L.,~Planck~TT,~TE,~EE~+~lowE~+~lensing~+~BAO})
\end{align}
respectively.

From the phenomenology of our model, reviewed briefly above, and discussed in detail in \cite{Dorlis:2024uei}, one can deduce:
\begin{align}\label{eiperiodicval}
\epsilon_1 \simeq 0.005\, , \quad \epsilon_2 &\simeq 0.02 +  14.6 \, \Big(\frac{M_s}{M_{\rm Pl}}\Big)^{-4} \, \frac{\Lambda_1^4}{|b(0)\, \Lambda_{\rm cond}^3|} \simeq  0.02 + 1.9 \times 10^{7} \, \Big(\frac{\Lambda_1}{M_s}\Big)^4 \nonumber \\
& \simeq \,  0.02 + 1.9 \, \xi^4 \, \times 10^{7} \, e^{-4\, \mathcal S_{\rm inst}} \lesssim   0.02 + 1.9 \, \xi^4 \, \times 10^{7} \, \exp\Big(-\frac{32\pi^2}{g^2_{\rm YM}}\Big)\,.
\end{align} 
Agreement with \eqref{nsplanck}
requires that the non-perturbative contributions to the $\epsilon_2$ are of order $\mathcal O(10^{-3})$, that is, one should impose:
\begin{align}\label{nonpertns}
\Delta n_s^{\rm non-perturb.} = - 1.9 \, \xi^4 \, \times 10^{7} \, e^{-4\, \mathcal S_{\rm inst}} = - \mathcal O(10^{-3}) \,  ,
\end{align}
which is achieved for gauge-sector (Euclidean) instanton actions of  order of magnitude (assuming for concreteness $\xi = \mathcal O(1)$) :
\begin{align}\label{instactionval}
\mathcal S_{\rm inst} \sim 5.9 \,.
\end{align} 
The latter would satisfy the bound \eqref{instbound} if  $5.9 \gtrsim 8\pi^2/g_{\rm YM}^2 $, corresponding to a fine structure constant $\alpha_{\rm YM}$ of the underlying Yang-Mills gauge theory (renormalized at the energy scale of the instantons)  $\alpha_{\rm YM} \equiv \frac{g_{\rm YM}^2}{4\pi} \gtrsim \frac{2\pi}{5.9} \sim 1.06$,
which is a natural value for a strongly coupled gauge theory. 

We complete our discussion by remarking that for the scale hierarchy  $\Lambda_1 \ll \Lambda_{\rm cond}$, which is the case we assume for our stringy RVM~\cite{Dorlis:2024uei}, the conclusions of the dynamical system analysis of \cite{Dorlis:2024yqw} regarding inflation and its duration~\cite{Dorlis:2024uei} remain valid in order of magnitude. One may generalise such  analysis by treating $\Lambda_{\rm cond}, \Lambda_1$ as arbitrary phenomenological parameters, and also including a second (compactification) axion. This will be the topic of forthcoming publications.

\section*{Acknowledgements}

N.E.M. wishes to thank the organisers of the Corfu2024 Workshop on the {\it Standard Model and Beyond} for the invitation to speak. The work of N.E.M. is supported in part by the UK Science and Technology Facilities research Council (STFC)  under the research grant ST/X000753/1. The
work of P.D. is supported by a graduate scholarship from the National Technical University of Athens (Greece).
The work of S.-N.V. is supported by the Hellenic Foundation for Research and Innovation
(H.F.R.I. (EL.ID.EK.)) under the “5th Call for H.F.R.I. Scholarships to PhD Candidates” (Scholarship Number:
20572). NEM also acknowledges participation in the COST Association Actions CA21136 “Addressing observational
tensions in cosmology with systematics and fundamental physics (CosmoVerse)” and CA23130 ”Bridging high and
low energies in search of quantum gravity (BridgeQG)”.

\bibliographystyle{JHEP}

\bibliography{bibliographyDMV_Corfu_April_2025}

\end{document}